\documentclass[12pt,preprint]{aastex}

\shorttitle{Interstellar Chlorine Abundance}
\shortauthors{Moomey et al.}

\begin{document}

\title{Revisiting the Chlorine Abundance in Diffuse Interstellar Clouds from 
Measurements with the $Copernicus$ Satellite}

\author{Daniel Moomey\altaffilmark{1}, S. R. Federman\altaffilmark{1},
and Y. Sheffer\altaffilmark{1}$^,$\altaffilmark{2}}

\altaffiltext{1}{Department of Physics and Astronomy, University of Toledo,
Toledo, OH 43606, USA; steven.federman@utoledo.edu}
\altaffiltext{2}{Department of Astronomy, University of Maryland, 
College Park, MD 20742, USA; ysheffer@astro.umd.edu}

\begin{abstract}
We reanalyzed interstellar Cl~{\small I} and Cl~{\small II} spectra 
acquired with the $Copernicus$ satellite.  The directions for this study 
come from those of Crenny \& Federman and sample the transition from atomic 
to molecular rich clouds where the unique chemistry leading to molecules 
containing chlorine is initiated.  Our profile syntheses relied on up-to-date 
laboratory oscillator strengths and component structures derived 
from published high-resolution measurements of K~{\small I} absorption that 
were supplemented with Ca~{\small II} and Na~{\small I} D results.  
We obtain self-consistent results for the Cl~{\small I} lines at 1088, 1097, 
and 1347 \AA\ from which precise column densities are derived.  The improved 
set of results reveals clearer correspondences with H$_2$ and 
total hydrogen column densities.  These linear relationships arise from 
rapid conversion of Cl$^+$ to Cl$^0$ in regions where H$_2$ is present.
\end{abstract}


\section{Introduction}

There is renewed interest in the elemental abundance of chlorine in 
diffuse interstellar clouds.  Previous observations (Jura \& York 1978; 
Harris \& Bromage 1984; Jenkins et al. 1986) focused on the unique 
chemical property that, of the abundant elements, only the dominant ion 
of Cl, Cl$^+$, has exothermic reactions with H$_2$, and 
equally important, they occur on nearly every collision.  As a 
result, observations acquired with the $Copernicus$ satellite and the 
{\it International Ultraviolet Explorer} ($IUE$) examined the relative 
amounts of Cl$^{\rm o}$ and Cl$^+$ as a function of H$_2$ column, 
average line-of-sight density, etc.  Jura \& York (1978) examined 
$Copernicus$ spectra for 10 sight lines and found that the interstellar 
chlorine abundance appeared to be relatively constant and that the 
results were consistent with their chemical model.  This work was expanded 
to 40 directions by Harris \& Bromage (1984), who relied on data for 
Cl~{\small I} $\lambda\lambda$1097, 1347 and Cl~{\small II} $\lambda$1071 
acquired with the $Copernicus$ satellite and compiled by Bohlin et al. 
(1983).  These were supplemented with their own spectra of $\lambda$1347 
obtained with $IUE$.  Harris \& Bromage surmised that the chlorine 
abundance did not vary appreciably with distance to the background star nor 
with $E$($B-V$), but that the abundance decreased with increasing average 
gas density, $<n>$ [$N_{tot}$(H)/distance to star, where $N_{tot}$(H) 
$=$ $N$(H~{\small I}) $+$ 2 $N$(H$_2$)].  Their expanded sample continued 
to support the chemical model.  As part of an analysis of interstellar 
abundances for several elements based on the data collected by Bohlin 
et al. (1983), Jenkins et al. (1986) found that the chlorine abundance 
varied with $<n>$ through the use of detailed statistical arguments.  

The more recent interest revolves around four issues.  
First, Sonnentrucker et al. (2002) 
described an analysis where the fraction of Cl in the neutral form 
provides information about the molecular hydrogen content of the gas.  
Second, detections of interstellar fluorine with the {\it Far Ultraviolet 
Spectroscopic Explorer} ($FUSE$) by Federman et al. (2005) and Snow, 
Destree, \& Jensen (2007) were interpreted in part through comparison 
with the Cl abundance because both halogens are expected to have similar 
levels of depletion onto grains.  Third, a self-consistent set of 
oscillator strengths ($f$-values) for lines of Cl~{\small I} 
($\lambda\lambda$1088, 1097, and 1347) and Cl~{\small II} ($\lambda$1071) is 
now available based on laboratory experiments (Schectman et al. 1993; 
2005) and theoretical calculations (Bi\'{e}mont, Gebarowski, \& 
Zeippen 1994; Tayal 2004; Froese Fischer, Tachiev, \& Irimia 2006; 
Oliver \& Hibbert 2007).  Fourth, measurements with the {\it Herschel 
Space Observatory} (Lis et al. 2010) of H$_2$Cl$^+$ and HCl in absorption 
from diffuse molecular gas allows a more complete analysis of chlorine 
chemistry.

We used the improved set of oscillator strengths to reevaluate the 
chemical trends involving neutral and singly ionized chlorine with 
atomic and molecular hydrogen.  The chlorine data come from the sight 
lines analyzed by Crenny \& Federman (2004) for CO absorption.  Since the 
$f$-value for the Cl~{\small II} line at 1071 \AA\ has not changed 
significantly from the value adopted by Jenkins et al. (1986), as noted by 
Schectman et al. (2005), we do not discuss the chlorine 
depletion patterns here.  Instead, the use of the three Cl~{\small I} 
lines, along with profile synthesis based on component structure for the 
line of sight from high-resolution ground-based measurements, provides 
the means to study the correspondences between the chlorine and hydrogen 
species with greater precision than was possible in the earlier work.

\section{Observations}

Earlier studies of Cl~{\small I} derived column densities either by 
assuming an optically thin line ($\lambda$1097) or through curves of 
growth that yielded an effective $b$-value consistent with the measured 
equivalent width ($W_{\lambda}$) for the line at 1347 \AA.  An 
effective $b$-value arises when the component structure from individual 
clouds is not resolved.  Our sample mainly comprises the directions from
Crenny \& Federman (2004) because Cl~{\small I} $\lambda$1088 is blended 
with the CO $C-X$ (0,0) band, which was the focus in that paper.  Once the 
column density of CO was known, the absorption from the $C-X$ (0,0) band 
could be removed from the spectrum, leaving only $\lambda$1088.  The 
importance of this line lies in the fact that it has a strength 
intermediate between those for the lines at 1097 and 1347 \AA.  To this 
sample, we added the observations on $\rho$ Oph A and $\chi$ Oph, which 
were analyzed for CO absorption by Federman et al. (2003).  Since 
chlorine depletion was not our focus, we did not include moderately 
reddened, molecule-rich directions such as $o$ Per, $\zeta$ Per, and 
$\zeta$ Oph.

High-resolution spectra acquired with the U1 photomultiplier tube were 
extracted from the $Copernicus$ archive at the Multiwavelength Archive 
at the Space Telescope Science Institute (MAST).  The spectra, with a 
nominal resolution of 0.05 \AA, revealed absorption from Cl~{\small II} 
$\lambda$1071 and the three lines of Cl~{\small I} at 1088, 1097, and 
1347 \AA.  The processing of the MAST data was essentially the same 
as that done on the CO data by Crenny \& Federman (2004).  Scans 
indicating stellar continua on both sides of the interstellar feature 
and free of peculiarities were rebinned and summed, resulting in a final 
spectrum for further analysis.  Using standard routines in the IRAF 
environment, rectified spectra were obtained through low-order 
polynomial fits to the stellar continua.  As with the data analyzed 
by Crenny \& Federman (2004), all MAST spectra were corrected for 
particle background, but stray light and scattered light were not 
removed because accidental blockage of the stray-light source may 
have occurred and scattered light could represent only 5 to 10\% of 
the local continuum.  The consequences of not accounting for these 
effects can be assessed by comparison with results from earlier 
studies.  This comparison of equivalent widths appears 
in Table 1.  Our uncertainties were determined by multiplying the 
rms deviations in the rectified stellar continuum by the width of the 
interstellar feature at the 50\% level.  When no line was present, 
3-$\sigma$ upper limits were derived.  

Our values of $W_{\lambda}$ agree well with previous determinations (Jura \& 
York 1978;  Bohlin et al. 1983; Federman 1986) when the mutual uncertainties 
are considered.  The occasional discrepancy involves measurements of 
$W_{1088}$ between Federman (1986) and us, such as toward $\epsilon$ 
Per and $\pi$ Sco.  The differences can be ascribed to the difficulty in 
accounting for absorption from the CO $C$ -- $X$ (0,0) band; however, for 
most sight lines the two sets of results agree.  An additional comparison
of $Copernicus$ data for $\lambda$1088 is possible; toward $\gamma$ Ara 
Morton \& Hu (1975) found $6.8\pm0.7$ m\AA\ versus our measurement of 
$6.3\pm0.4$ m\AA.  Finally, the line at 1347 \AA\ was observed toward several 
bright stars with grating G160M of the Goddard High Resolution Spectrograph 
on the {\it Hubble Space Telescope}: $\eta$ Tau ($21.8\pm0.7$ vs. 
$15.4\pm3.1$ m\AA), 1 Sco ($27.9\pm0.9$ vs. $20.5\pm4.8$ m\AA), $\pi$ Sco 
($15.8\pm0.4$ vs. $15.5\pm1.3$ m\AA), and $\sigma$ Sco ($21.2\pm0.4$ vs. 
$21.4\pm4.3$ m\AA).  The correspondence is very good, within the 2-$\sigma$ 
uncertainties of the $Copernicus$ results in all cases.  We believe, 
therefore, that the effects of stray light and scattered light do not 
impact our results in a significant way.

\section{Results and Discussion}

A further refinement in our analysis is the use of profile synthesis to 
extract column densities, taking into account known component structures.  
As in much of our previous syntheses, we utilized the fitting routine 
ISMOD (Sheffer, unpublished) with the laboratory $f$-values reported 
by Schectman et al. (1993, 2005).  The component structure was inferred 
from published ground-based observations of  K~{\small I} (Welty \& Hobbs 
2001; Pan et al.  2004), Ca~{\small II} (Vallerga et al. 1993; Welty, Morton, 
\& Hobbs 1996; Welsh et al. 1997; Pan et al. 2004; Ritchey et al. 2006), 
and Na~{\small I} D (Crawford, Barlow, \& Blades 1989; Welsh et al. 
1991; Welty, Hobbs, \& Kulkarni 1994; White et al. 2001), all of which 
were acquired at spectral resolution much higher than that of the 
$Copernicus$ measurements.  Because the Na~{\small I} D lines are 
usually severely optically thick, even in very diffuse sight lines, we 
treated these measurements as secondary.  The first step was to check for 
consistency in component structure among independent studies for a specific 
species and from one species to the other.  The correspondence was quite 
good.  For the most part, we relied on the components seen in K~{\small I} 
for velocities, but allowed the $b$-values and fractional column densities 
to vary.  Occasionally, we had to combine nearby components, weighted 
according to component column density, in order to not 
overfit the observed absorption; this was necessary for the sight lines 
toward 139 Tau, 1 Sco, $\pi$ Sco, $\nu$ Sco, $\chi$ Oph, and 59 Cyg.  In a 
few cases, such as $\gamma$ Ara, we did not consider absorption from a 
substantially weaker component.  For HD~21278, $\alpha$ Cam, and 1 Cas, 
high-resolution component structure was not available, and so we fit 
absorption in the latter two directions with one or two components having 
$b$-values less than about 2 km s$^{-1}$ (see e.g., Pan et al. 2005 for 
discussion of limits).  For HD~21278, incorrect removal of CO absorption 
in the vicinity of $\lambda$1088 and the lack of Cl~{\small II} prevented 
us from deriving reliable results; this sight line is not considered in 
the analyses presented below. 

Final column densities [$N$(X)] were derived through the following steps.  
First, the Cl~{\small I} lines were fit with the same component structure 
from the ground-based spectra.  If any of the lines were overfit (as 
revealed by significantly less noise in the residuals within the range 
of absorption compared to the continuum), fits with fewer components were 
attempted.  The individual uncertainties in column density were inferred 
from the uncertainty in $W_{\lambda}$ for the line.  
Once consistent column densities were obtained for the 
Cl~{\small I} lines, a weighted average yielded our preferred value for 
$N$(Cl~{\small I}).  The same component structure was then used to fit 
the absorption from Cl~{\small II} $\lambda$1071.  Figure 1 shows the 
results of the profile synthesis for the gas toward $\epsilon$ Per and 
$\sigma$ Sco.  The spectral ranges for $\lambda\lambda$1088, 1097 shown 
for $\sigma$ Sco were limited by the extent of the original scans.

Our column densities, along with those of earlier analyses of 
$Copernicus$ spectra (Jura \& York 1978; Harris \& Bromage 1984; Jenkins 
et al. 1986) scaled to the oscillator strengths used here, appear in 
Table 2, and the values of $W_{\lambda}$ from the fits are given in Table 1.  
Our measured $W_{\lambda}$s and the fitted values 
are in close agreement.  As for the comparison 
of column densities, we first note that the results of Harris \& Bromage 
(1984) in Table 2 were based in part on analysis of $IUE$ spectra, 
which were not considered by us because they have 
lower spectral resolution and generally lower signal to noise.  Moreover, 
the results of Jenkins et al. (1986) are based on the data compiled in 
Bohlin et al. (1983).  Previous and our determinations agree for the most 
part, with ours having improved precision.  This arises from our use of 
more than one line for Cl~{\small I} and from the adoption of component 
structure from high-resolution spectra of K~{\small I}, Ca~{\small II}, 
and Na~{\small I} D.  It is also worth comparing the $b$-values deduced 
by Jenkins et al. (1986) with our component structure.  For the sight 
lines with a single component (1 Sco, $\nu$ Sco, $\mu$ Nor, 59 Cyg, and 
1 Cas), the $b$-values are very similar.  For the other sight lines, 
the $b$-values from Jenkins et al. (1986) are larger than ours, but 
instead are consistent with the range in velocities for the components 
used in our study.  In essence, the values quoted by Jenkins et al. 
(1986) for the latter group are effective $b$-values.

We examined the resulting trends among $N$(Cl~{\small I}) and 
$N$(Cl~{\small II}) with $N$(H~{\small I}) and $N$(H$_2$), as well as 
$N_{tot}$(Cl) [$N$(Cl~{\small I}) $+$ $N$(Cl~{\small II})] versus 
$N_{tot}$(H).  The hydrogen results come from Savage et al. (1977).  
For the sight lines in common, the values of $N$(H~{\small I}) 
of Savage et al. agree with those of Diplas \& 
Savage (1994), considering the uncertainties quoted ($\approx$ 0.10 dex) 
by the latter authors.  We, therefore, assigned uncertainties of 0.10 dex 
to $N$(H~{\small I}) for our sample.

Only two relationships revealed significant trends, $N$(Cl~{\small I}) 
vs. $N$(H$_2$) and $N_{tot}$(Cl) vs. $N_{tot}$(H), and they are shown 
in Figures 2 and 3.  [These two relationships had correlation coefficients 
($r^2$) greater than 0.6, compared to $\le$0.3 
for the others.]  Our BCES least-squares 
fits (Akritas \& Bershady 1996), which allow the uncertainties in x and y 
to be treated independently, are also given.  In Fig. 3, directions with 
limits on $N$(Cl~{\small II}) or those without H~{\small I} column 
densities are not included in the fits.  The latter three points lie at 
least 1.0 dex to the left of the data shown here.

Figure 2 shows the relationship between neutral chlorine and 
molecular hydrogen.  The data reveal a linear trend with a slope of 
$0.92\pm0.19$ that spans a range in column densities of nearly 100.  
The close correspondence between Cl$^0$ and H$_2$ arises from 
rapid hydrogen abstraction reactions between Cl$^+$ and H$_2$ and the 
subsequent dissociative recombination of the molecular ions and the 
photodissociation of HCl (e.g., Jura 1974; Lis et al. 2010).  This figure 
provides a firm foundation for deriving the molecular hydrogen content 
from observations of neutral chlorine (Sonnentrucker et al. 2002).

Next we consider the trend seen in Figure 3, a plot of log~$N_{tot}$(Cl) vs. 
log~$N_{tot}$(H).  The BCES fit again suggests a linear relationship, 
with a slope of $1.07\pm0.14$.  Most of the sight lines with upper 
limits to $N$(Cl~{\small II}) -- $\alpha$ Cam, 59 Cyg, and 1 Cas -- are 
consistent with the fit, as is the result for 67 Oph, which has no 
spectrum of Cl~{\small II} $\lambda$1071.  Only the diffuse gas toward 
$\nu$ Sco suggests the presence of an anomaly, but we note that the 
limited spectral coverage in the vicinity of $\lambda$1071 might have 
compromised the result.  Of the directions in our sample, only the 
sight line toward $\rho$ Oph indicates depletion of chlorine onto 
grains.  This is not unexpected because it has the largest $<n>$.  
Thus, $\rho$ Oph was excluded from the fit as well.  A weighted 
average of the results for directions indicated by 
filled circles yields a chlorine abundance, $N_{tot}$(Cl)/$N_{tot}$(H), 
of $-6.99\pm0.04$.  Our inferred abundance compares favorably with 
previous determinations: $-6.95\pm0.03$ (Harris \& Bromage 1984) and 
$-7.09\pm0.03$ (Jenkins et al. 1986).  Both previous values have been 
scaled to reflect the revised $f$-value given by Schectman et al. (2005) 
for Cl~{\small II}, whose absorption dominates the low-density sight 
lines considered by Harris \& Bromage and Jenkins et al.  When compared 
with the meteoritic abundance of $-6.74\pm0.06$ (Lodders 2003), the 
interstellar gas phase abundance of chlorine indicates a factor of nearly 
two depletion onto grains.

At first glance, it may seem surprising that our sample of directions, 
which probe the transition from atomic to molecular gas, shows such 
a correspondence between $N_{tot}$(Cl) and $N_{tot}$(H) and a chlorine 
abundance indistinguishable from earlier determinations.  Our sample 
includes some sight lines where most of the Cl is singly-ionized, while 
others are dominated by neutral chlorine.  Moreover, the average densities 
for the gas toward our targets range between $<n>$ of 0.25 cm$^{-3}$ and 
6 cm$^{-3}$, except for gas toward $\rho$ Oph, which has $<n>$ greater than 
10 cm$^{-3}$.  Usually, interstellar abundances are inferred from results 
for sight lines with $<n>$ less than about 0.1 cm$^{-3}$ (e.g., Jenkins 
et al. 1986).  We suggest that the difference with results for other 
elements discussed by Jenkins et al. (1986) -- Mg, P, Mn, and Fe -- and 
by Ritchey et al. (2011) -- B, O, Cu, and Ga, for instance, 
again is connected to the rapid reactions between Cl$^+$ and 
H$_2$, resulting in conversion of Cl$^+$ to Cl$^0$ that occurs on time scales 
that are short compared to those associated with depletion onto grains.  
Based on the rate coefficient for Cl$^+$ $+$ H$_2$ $\rightarrow$ HCl$^+$ 
$+$ H, the initial reaction in the sequence, given by Anicich (1993) 
and a density of 1 cm$^{-3}$, we obtain a rate of about $10^{-9}$ s$^{-1}$, 
or a time scale of about 30 years.  Other reactions in the sequence are 
equally fast.  We also note that the range in 
$N$(Cl~{\small II})/$N$(Cl~{\small I}) ratios found for our sample is the 
likely cause for the weak correlation between Cl$^+$ and atomic H.

As noted in the introduction, we undertook this study in part because 
a self-consistent set of $f$-values was emerging for the Cl~{\small I} 
lines commonly studied in diffuse clouds, $\lambda\lambda$1088, 1097, 
and 1347.  This came about in part through the realization that energy 
levels required new identifications (Bi\'{e}mont et al. 1994; Oliver \& 
Hibbert 2007).  However, other lines are seen in interstellar spectra.  
Sonnentrucker, Friedman, \& York (2006) used the results of Schectman et al. 
(1993) as a starting point in deriving $f$-values for relatively weak 
Cl~{\small I} lines at 1004, 1079, 1090, and 1094 \AA\ seen in spectra 
acquired with $FUSE$.  These lines are especially useful for more reddened 
directions.  Here, the correspondence between empirical measures and 
theoretical calculations of Bi\'{e}mont et al. (1994) is less satisfactory, 
while that for the results of Oliver \& Hibbert (2007) for $\lambda$1094 
is good.  Their reevaluation of assigning term values 
to a number of energy levels probably led to the improvement.  

Lis et al. (2010) detected absorption from H$_2$Cl$^+$ and HCl in 
diffuse molecular clouds along the line of sight toward the star-forming 
region Sgr~B2(S).  They derived a combined abundance for these 
chlorine molecules of about $4 \times 10^{-9}$, which is about 1\% of the 
total chlorine budget (see above).  The remainder is likely in atomic 
forms and depleted onto interstellar grains.  This is consistent with 
the tentative detection of HCl absorption at UV wavelengths seen toward 
$\zeta$ Oph (Federman et al. 1995).  Furthermore, the HCl/H$_2$Cl$^+$ ratio 
derived by Lis et al. (2010), about 1, agrees with expectations from modeling 
diffuse clouds.

\section{Conclusions}

We reanalyzed a sample of directions that were observed with the $Copernicus$ 
satellite and that involve diffuse clouds where the transition from 
atomic to molecular gas takes place.  The basis for the sample comes from 
the work of Crenny \& Federman (2004), who studied CO absorption, but our 
focus was on data for lines of Cl~{\small I} and Cl~{\small II}.  Column 
densities were extracted from profile syntheses that relied on up-to-date 
oscillator strengths and component structures from higher resolution, 
ground-based observations.  The result is the most precisely determined 
set of column densities for our sample to date.  We examined the 
correspondence between $N$({Cl~\small I}) and $N$(H$_2$) and between 
$N_{tot}$(Cl) and $N_{tot}$(H) and found strong linear trends in both cases.  
We attributed the trends to the rare instance of a rapid ion-molecule 
reaction between a singly ionized atom, Cl$^+$, and H$_2$.  Our findings 
are likely to aid in interpreting observations of HCl and H$_2$Cl$^+$ 
absorption from diffuse clouds revealed by $Herschel$.

\acknowledgments
This work was supported by NASA grant NNG 06-GC70G.  We acknowledge the 
assistance provided by Michael Stone in the early phases of the research 
reported here and the helpful exchanges with Don York, Ed Jenkins, and 
Jim Lauroesch regarding the background in $Copernicus$ spectra.  
We utilized the $Copernicus$ archive available at the Multiwavelength
Archive at the Space Telescope Science Institute.  Additional
observations made with the NASA/ESA Hubble Space Telescope were obtained 
from the data archive at STScI.  STScI is operated by the Association 
of Universities for Research in Astronomy, Inc. under NASA contract 
NAS5-26555.

\clearpage

\begin{deluxetable}{llcccccc}
\tablecaption{Comparison of Equivalent Widths (in m\AA)}
\tablecolumns{8}
\tablewidth{0pt}
\tabletypesize{\scriptsize}
\tablehead{
\colhead{Star} & \colhead{Name} & \colhead{Line (\AA)\tablenotemark{a}} 
& \colhead{This Work} & \colhead{JY\tablenotemark{b}} 
& \colhead{BJSYHSS\tablenotemark{c}} & \colhead{F\tablenotemark{d}} & 
\colhead{Fit}}
\startdata
HD 23408 & 20 Tau &1071 &  $\le$6.8 & $\ldots$ & 15.0$\pm$15.3 & 
$\ldots$ & $\le$2.0 \\
 & & 1088 & 23.5$\pm$5.2 & $\ldots$ & $\ldots$ & $\ldots$ & 14.9 \\
 & & 1097 & $\le$35.6 & $\ldots$ & $\ldots$ & $\ldots$ & 3.0 \\
 & & 1347 & 22.2$\pm$10.3 & $\ldots$ & 11.3$\pm$15.1 & $\ldots$ & 24.5 \\
HD 23630 & $\eta$ Tau & 1071 & $\le$6.9 & $\ldots$ & $-$3.0$\pm$18.3 & 
$\ldots$ & $\le$2.1 \\
 & & 1088 & $\le$4.3 & $\ldots$ & $\ldots$ & 3$\pm$2 & 11.3 \\
 & & 1097 & 3.7$\pm$1.7 & $\ldots$ & $\ldots$ & $\ldots$ & 2.8 \\
 & & 1347 & 15.4$\pm$3.1 & $\ldots$ & 25.4$\pm$7.4 & $\ldots$ & 17.6 \\
HD 24760 & $\epsilon$ Per & 1071 & 2.7$\pm$0.7 & 3.0$\pm$1 & 2.3$\pm$0.6 & 
$\ldots$ & 2.8 \\
 & & 1088 & 8.1$\pm$0.4	& $\ldots$ & $\ldots$ & 15$\pm$2 & 7.9 \\
 & & 1097 & 0.9$\pm$0.2	& 2.5$\pm$0.7 & $\ldots$ & $\ldots$ & 1.0 \\
 & & 1347 & 21.4$\pm$0.6 & 22$\pm$2 & 22.5$\pm$1.4 & $\ldots$ & 18.9  \\
HD 30614 & $\alpha$ Cam	& 1071 & $\le$3.6 & 15$\pm$6 & 11.4$\pm$4.9 & 
$\ldots$ & $\le$1.0 \\
 & & 1088 & 37.4$\pm$3.1 & $\ldots$ & $\ldots$ & 23$\pm$2 & 37.0 \\
 & & 1097 & 9.4$\pm$1.8	& 11$\pm$4 & 16.2$\pm$3.1 & $\ldots$ & 10.3 \\
HD 36861 & $\lambda$ Ori & 1071 & 5.9$\pm$1.3 & 7$\pm$1	& 7.8$\pm$1.0 & 
$\ldots$ & 6.0 \\
 & & 1088 & 10.9$\pm$0.9 & $\ldots$ & $\ldots$ & 10$\pm$1 & 11.2 \\
 & & 1097 & 3.0$\pm$0.9	& 3$\pm$1 & $\ldots$ & $\ldots$ & 1.7 \\
HD 40111 & 139 Tau & 1071 & 8.4$\pm$2.4	& $\ldots$ & 12.5$\pm$2.3 & 
$\ldots$ & 10.2 \\
 & & 1088 & 8.2$\pm$1.0 & $\ldots$ & $\ldots$ & $\ldots$ & 11.4 \\
 & & 1097 & $\le$2.6 & $\ldots$ & 1.4$\pm$2.0 & $\ldots$ & 1.4 \\
 & & 1347 & 27.8$\pm$2.2 & $\ldots$ & 33.3$\pm$5.3 & $\ldots$ & 28.8 \\
HD 141637 & 1 Sco & 1071 & 9.0$\pm$4.4 & $\ldots$ & 17.3$\pm$5.9 & 
$\ldots$ & 9.6 \\
 & & 1088 & 5.4$\pm$1.2 & $\ldots$ & $\ldots$ & $\ldots$ & 7.5 \\
 & & 1097 & $\le$3.1 & $\ldots$ & 3.0$\pm$2.0 & $\ldots$ & 0.9 \\
 & & 1347 & 20.5$\pm$4.8 & $\ldots$ & 29.6$\pm$6.3 & $\ldots$ & 18.5 \\
HD 143018 & $\pi$ Sco & 1071 & 7.4$\pm$2.2 & $\ldots$ & $\ldots$ & 
$\ldots$ & 6.4 \\
 & & 1088 & 4.8$\pm$0.2	& $\ldots$ & $\ldots$ & 8$\pm$1 & 5.2 \\
 & & 1097 & $\le$0.5 & $\ldots$ & $\ldots$ & $\ldots$ & 0.6 \\
 & & 1347 & 15.5$\pm$1.3 & $\ldots$ & $\ldots$ & $\ldots$ & 13.0 \\ 
HD 143275 & $\delta$ Sco & 1071 & 10.8$\pm$0.2 & $\ldots$ & 10.9$\pm$0.4 & 
$\ldots$ & 10.1 \\
 & & 1088 & 16.1$\pm$1.0 & $\ldots$ & $\ldots$ & 18$\pm$1 & 15.4 \\
 & & 1097 & 4.2$\pm$0.3	& $\ldots$ & 3.1$\pm$0.4 & $\ldots$ & 3.1 \\
HD 144217 & $\beta^1$ Sco & 1071 & 8.4$\pm$0.2 & $\ldots$ & 9.8$\pm$0.8 & 
$\ldots$ & 9.0 \\
 & & 1088 & 18.9$\pm$0.4 & $\ldots$ & $\ldots$ & $\ldots$ & 19.0 \\
 & & 1097 & 4.4$\pm$0.4	& $\ldots$ & 4.9$\pm$0.7 & $\ldots$ & 5.1 \\
HD 144470 & $\omega^1$ Sco & 1071 & 14.9$\pm$3.9 & $\ldots$ & 16.3$\pm$1.8 & 
$\ldots$ & 14.8 \\
 & & 1088 & 17.7$\pm$1.4 & $\ldots$ & $\ldots$ & 23$\pm$6 & 18.0 \\
 & & 1097 & 3.0$\pm$1.0	& $\ldots$ & 3.4$\pm$1.7 & $\ldots$ & 3.6 \\
HD 145502 & $\nu$ Sco & 1071 & $\le$6.3	& $\ldots$ & 8.1$\pm$7.6 & 
$\ldots$ & $\le$1.4 \\
 & & 1088 & 10.5$\pm$0.9 & $\ldots$ & $\ldots$ & 15$\pm$2 & 12.8 \\
 & & 1097 & $\le$7.4 & $\ldots$ & 13.7$\pm$8.0 & $\ldots$ & 1.8 \\
 & & 1347 & 26.9$\pm$6.1 & $\ldots$ & 36.5$\pm$14.8 & $\ldots$ & 27.0 \\
 \\
 \\
 \\
HD 147165 & $\sigma$ Sco & 1071 & 21.7$\pm$4.2 & $\ldots$ & 21.5$\pm$2.6 & 
$\ldots$ & 20.4 \\
 & & 1088 & 11.9$\pm$0.5 & $\ldots$ & $\ldots$ & $\ldots$ & 12.0 \\
 & & 1097 & 0.7$\pm$0.3	& $\ldots$ & 7.9$\pm$2.9 & $\ldots$ & 1.7 \\
 & & 1347 & 21.4$\pm$2.0 & $\ldots$ & 23.1$\pm$2.9 & $\ldots$ & 25.4 \\
HD 147933 & $\rho$ Oph A & 1071 & 7.1$\pm$2.3 & $\ldots$ & 15.6$\pm$4.6 & 
$\ldots$ & 7.1 \\
 & & 1088 & 17.2$\pm$2.4 & $\ldots$ & $\ldots$ & 29$\pm$7 & 14.8 \\
 & & 1097 & 3.1$\pm$0.8	& $\ldots$ & 8.3$\pm$4.0 & $\ldots$ & 3.0 \\
 & & 1347 & 20.5$\pm$4.3 & $\ldots$ & 22.7$\pm$9.1 & $\ldots$ & 24.9 \\
HD 148184 & $\chi$ Oph & 1071 & 5.3$\pm$1.1 & $\ldots$ & 8.3$\pm$4.5 & 
$\ldots$ & 5.5 \\
 & & 1088 & 16.1$\pm$2.5 & $\ldots$ & $\ldots$ & $\ldots$ & 26.0 \\
 & & 1097 & 10.9$\pm$2.3 & $\ldots$ & 14.4$\pm$3.6 & $\ldots$ & 11.4 \\
 & & 1347 & 39.2$\pm$7.0 & $\ldots$ & 42.4$\pm$6.6 & $\ldots$ & 37.4 \\
HD 149038 & $\mu$ Nor & 1071 & 8.4$\pm$3.0 & $\ldots$ & 61.6$\pm$25.6 & 
$\ldots$ & 8.0 \\
 & & 1088 & 33.0$\pm$2.4 & $\ldots$ & $\ldots$ & 36$^{+16}_{-9}$ & 34.0 \\
 & & 1097 & 20.4$\pm$3.6 & $\ldots$ & 3.5$\pm$15.8 & $\ldots$ & 12.2 \\
HD 157246 & $\gamma$ Ara & 1071 & 15.4$\pm$1.5 & 9$\pm$2 & 20.9$\pm$1.3 & 
$\ldots$ & 12.3 \\
 & & 1088 & 6.3$\pm$0.4	& $\ldots$ & $\ldots$ & $\ldots$ & 7.1 \\
 & & 1097 & 2.1$\pm$0.7	& 2.5$\pm$1.4 & $\ldots$ & $\ldots$ & 0.8 \\
 & & 1347 & 18.9$\pm$0.9 & 22$\pm$6 & 23.5$\pm$1.5 & $\ldots$ & 18.8 \\
HD 164353 & 67 Oph & 1088 & 23.6$\pm$3.5 & $\ldots$ & $\ldots$ & 
11$^{+7}_{-6}$ & 19.9 \\
HD 200120 & 59 Cyg & 1071 & $\le$3.1 & $\ldots$ & 2.7$\pm$2.8 & 
$\ldots$ & $\le$1.9 \\
 & & 1088 & 2.5$\pm$0.9	& $\ldots$ & $\ldots$ & 6$\pm$2 & 3.4 \\
 & & 1097 & 4.4$\pm$1.6	& $\ldots$ & 0.8$\pm$2.6 & $\ldots$ & 0.5 \\
 & & 1347 & 8.3$\pm$3.2	& $\ldots$ & 15.9$\pm$4.6 & $\ldots$ & 7.7 \\
HD 217675 & $o$ And & 1071 & $\ldots$ & $\ldots$ & 1.0$\pm$9.4 & 
$\ldots$ & \tablenotemark{e} \\
 & & 1088 & 11.9$\pm$1.4 & $\ldots$ & $\ldots$ & 14$\pm$2 & 10.8 \\
 & & 1347 & 21.7$\pm$2.7 & $\ldots$ & 25.9$\pm$6.1 & $\ldots$ & 23.1 \\
HD 218376 & 1 Cas & 1071 & $\le$14.7 & $\ldots$ & 10.3$\pm$5.1 & 
$\ldots$ & $\le$6.6 \\
 & & 1088 & 18.3$\pm$9.6 & $\ldots$ & $\ldots$ & $\ldots$ & 21.9 \\
 & & 1097 & 6.5$\pm$2.0	& $\ldots$ & 12.7$\pm$4.5 & $\ldots$ & 7.2 \\
\enddata
\tablenotetext{a}{The entries refer to Cl {\tiny II} $\lambda1071$; 
Cl {\tiny I} $\lambda\lambda1088,1097,1347$.}
\tablenotetext{b}{Jura \& York 1978.}
\tablenotetext{c}{Bohlin et al. 1983.}
\tablenotetext{d}{Federman 1986.}
\tablenotetext{e}{This line is severely blended with absorption from H$_2$.}
\end{deluxetable}

\clearpage

\begin{deluxetable}{lcccccc}
\tablecaption{Column Densities}
\tablecolumns{7}
\tablewidth{0pt}
\tabletypesize{\scriptsize}
\tablehead{
\colhead{Star} & \colhead{$N$(Cl~{\tiny II})\tablenotemark{a}} & 
\colhead{$N$(Cl~{\tiny I})\tablenotemark{a}} & 
\colhead{$N_{tot}$(Cl)\tablenotemark{a}} & 
\colhead{log $N$(H~{\tiny I})\tablenotemark{b}} & 
\colhead{log $N$(H$_2$)\tablenotemark{b}} & 
\colhead{log $N_{tot}$(H)\tablenotemark{b}}}
\startdata
20 Tau & $\le$1.4 & 3.5$\pm$0.8 & $\le$4.9 & $\ldots$ & 19.75 & $\ge$20.05 \\
 & 11.0(0.0-129)\tablenotemark{c}$^,$\tablenotemark{d} &
0.9(0.5-3.7)\tablenotemark{c} & \\
$\eta$ Tau & $\le$1.5 & 3.4$\pm$0.6 & $\le$4.9 & $\ldots$ & 19.54 & 
$\ge$19.84 \\
 & & 1.3(0.4-sat.)\tablenotemark{c} & \\
$\epsilon$ Per & 1.9$\pm$0.5 & 1.1$\pm$0.1 & 3.0$\pm$0.5 & 20.40 & 19.53 & 
20.50 \\
 & 1.4(0.6-2.4)\tablenotemark{c} & 1.3(1.0-2.2)\tablenotemark{c} & \\
 & 1.9(1.2-2.3)\tablenotemark{e} & 1.8(1.2-3.0)\tablenotemark{e} & \\
 & 1.4(1.0-1.8)\tablenotemark{f} & 1.0(0.9-1.1)\tablenotemark{f} & \\
$\alpha$ Cam & $\le$2.5 & 13.0$\pm$1.0 & $\le$15.5 & 20.90 & 20.34 & 21.09 \\
 & 7.8(0.9-16.2)\tablenotemark{c} & 20.0(8.9-35.6)\tablenotemark{c} & \\
 & 9.3(5.9-11.7)\tablenotemark{e} & 11.8(7.4-18.7)\tablenotemark{e} & \\
 & 7.1(4.0-10.0)\tablenotemark{f} & 17.5(13.8-20.5)\tablenotemark{f} & \\
$\lambda$ Ori & 4.7$\pm$1.0 & 1.9$\pm$0.2 & 6.6$\pm$1.0 & 20.78 & 19.11 & 
20.80 \\
 & 5.1(3.7-6.9)\tablenotemark{c} & 1.2(0.9-1.6)\tablenotemark{c} & \\
 & 4.4(3.5-5.5)\tablenotemark{e} & 3.2(2.0-4.0)\tablenotemark{e} & \\
 & 4.8(4.2-5.5)\tablenotemark{f} & 1.1(1.0-1.2)\tablenotemark{f} & \\
139 Tau & 7.4$\pm$2.1 & 1.5$\pm$0.1 & 8.9$\pm$2.1 & 20.90 & 19.74 & 20.96 \\
 & 7.8(0.9-16.2)\tablenotemark{c} & 3.6(1.7-7.1)\tablenotemark{c} & \\
 & 7.8(6.3-9.1)\tablenotemark{f} & 1.6(1.3-2.0)\tablenotemark{f} & \\
1 Sco & 7.5$\pm$3.7 & 1.0$\pm$0.1 & 8.5$\pm$3.7 & 21.19 & 19.23 & 21.20 \\
 & 8.7(5.2-13.2)\tablenotemark{c} & 3.2(1.3-11.2)\tablenotemark{c} & \\
 & 10.7(7.1-14.4)\tablenotemark{f} & 1.4(1.0-1.8)\tablenotemark{f} & \\
$\pi$ Sco & 4.9$\pm$1.5 & 0.70$\pm$0.03 & 5.6$\pm$1.5 & 20.72 & 19.32 & 
20.75 \\
$\delta$ Sco & 9.9$\pm$0.2 & 3.7$\pm$0.2 & 13.6$\pm$0.3 & 21.15 & 19.41 & 
21.16 \\
 & 7.6(6.8-8.5)\tablenotemark{c} & 3.6(2.8-4.5)\tablenotemark{c} & \\
 & 6.8(6.5-7.1)\tablenotemark{f} & 3.3(2.9-3.7)\tablenotemark{f} & \\
$\beta^1$ Sco & 8.4$\pm$0.2 & 6.5$\pm$0.1 & 14.9$\pm$0.3 & 21.09 & 19.83 & 
21.14 \\
 & 6.6(5.5-7.9)\tablenotemark{c} & 6.3(4.5-8.0)\tablenotemark{c} & \\
 & 6.0(5.6-6.6)\tablenotemark{f} & 5.3(4.5-5.9)\tablenotemark{f} & \\
$\omega^1$ Sco & 16.0$\pm$4.0 & 4.2$\pm$0.3 & 20.2$\pm$4.0 & 21.18 & 20.05 & 
21.24 \\
 & 11.7(8.7-16.2)\tablenotemark{c} & 5.0(3.2-8.9)\tablenotemark{c} & \\
 & 10.0(8.9-11.2)\tablenotemark{f} & 3.6(1.8-5.4)\tablenotemark{f} & \\
$\nu$ Sco & $\le$1.0 & 2.0$\pm$0.2 & $\le$3.0 & 21.15 & 19.89 & 21.19 \\
 & 4.7(0.0-26.9)\tablenotemark{c} & 20.0(0.0-sat.)\tablenotemark{c} & \\
$\sigma$ Sco & 25.0$\pm$5.0 & 1.9$\pm$0.1 & 26.9$\pm$5.0 & 21.34 & 19.79 & 
21.37 \\
 & 17.8(12.0-28.2)\tablenotemark{c} & 11.2(0.8-sat.)\tablenotemark{c} & \\
 & 13.2(11.7-14.8)\tablenotemark{f} & 8.3(5.3-11.5)\tablenotemark{f} & \\
$\rho$ Oph A & 6.0$\pm$1.9 & 3.5$\pm$0.4 & 9.5$\pm$1.9 & 21.81 & 20.57 & 
21.86 \\
 & 6.6(5.5-7.9)\tablenotemark{c} & 6.3(4.5-8.0)\tablenotemark{c} & \\
 & 9.8(6.8-12.6)\tablenotemark{f} & 8.9(4.6-18.6)\tablenotemark{f} & \\
$\chi$ Oph & 4.2$\pm$0.9 & 18.0$\pm$1.0 & 22.2$\pm$1.2 & 21.15 & 20.63 & 
21.35 \\
 & 11.7(8.7-16.2)\tablenotemark{c} & 5.0(3.2-8.9)\tablenotemark{c} & \\
 & 5.1(2.3-7.9)\tablenotemark{f} & 15.2(11.5-19.1)\tablenotemark{f} & \\
$\mu$ Nor & 6.2$\pm$2.2 & 17.0$\pm$1.0 & 23.2$\pm$1.7 & 21.00 & 20.44 & 
21.19 \\
 & 5.8(0.0-26.9)\tablenotemark{c} & 20.0(0.8-sat.)\tablenotemark{c} & \\
 & 38.0(22.4-53.7)\tablenotemark{f} & 13.2(9.4-21.9)\tablenotemark{f} & \\
$\gamma$ Ara & 9.2$\pm$0.9 & 0.90$\pm$0.04 & 10.1$\pm$0.9 & 20.68 & 19.24 & 
20.71 \\
 & 18.2(14.1-22.9)\tablenotemark{c} & 1.3(1.0-1.6)\tablenotemark{c} & \\
 & 5.6(4.5-7.1)\tablenotemark{e} & 2.6(1.3-4.2)\tablenotemark{e} & \\
 & 12.9(12.0-13.8)\tablenotemark{f} & 1.1(1.0-1.2)\tablenotemark{f} & \\
67 Oph & $\ldots$ & 3.4$\pm$0.5 & $\ge$3.4 & 21.00 & 20.26 & 21.14 \\
59 Cyg & $\le$2.5 & 0.50$\pm$0.12 & $\le$3.0 & 20.26 & 19.32 & 20.34 \\
 & 1.7(0.0-6.5)\tablenotemark{c} & 1.4(0.4-sat.)\tablenotemark{c} & \\
$o$ And & $\ldots$ & 1.5$\pm$0.1 & $\ge$1.5 & $\ldots$ & 19.67 & $\ge$19.97 \\
 & & 1.8(0.6-sat.)\tablenotemark{c} & \\
1 Cas & $\le5.4$ & 9.6$\pm$2.6 & $\le$15.0 & 20.95 & 20.15& 21.07 \\
 & 2.2(0.0-5.4)\tablenotemark{c} & 7.1(4.0-14.1)\tablenotemark{c} & \\
 & 6.3(3.2-9.5)\tablenotemark{f} & 13.5(8.7-18.3)\tablenotemark{f} & \\
\enddata
\tablenotetext{a}{Present results given in first row; other 
determinations appear in subsequent rows.  The column densities have units 
of 10$^{13}$ cm$^{-2}$.}
\tablenotetext{b}{Hydrogen results from Savage et al. 1977.}
\tablenotetext{c}{Jenkins et al. 1986.}
\tablenotetext{d}{The values in parentheses indicate quoted range in column 
density.  `sat.' indicates upper limits greatly affected by optical depth.}
\tablenotetext{e}{Jura \& York 1978.}
\tablenotetext{f}{Harris \& Bromage 1984.}
\end{deluxetable}

\clearpage

\begin{figure}
\hspace{0.7in}
\includegraphics[scale=0.67, angle=90]{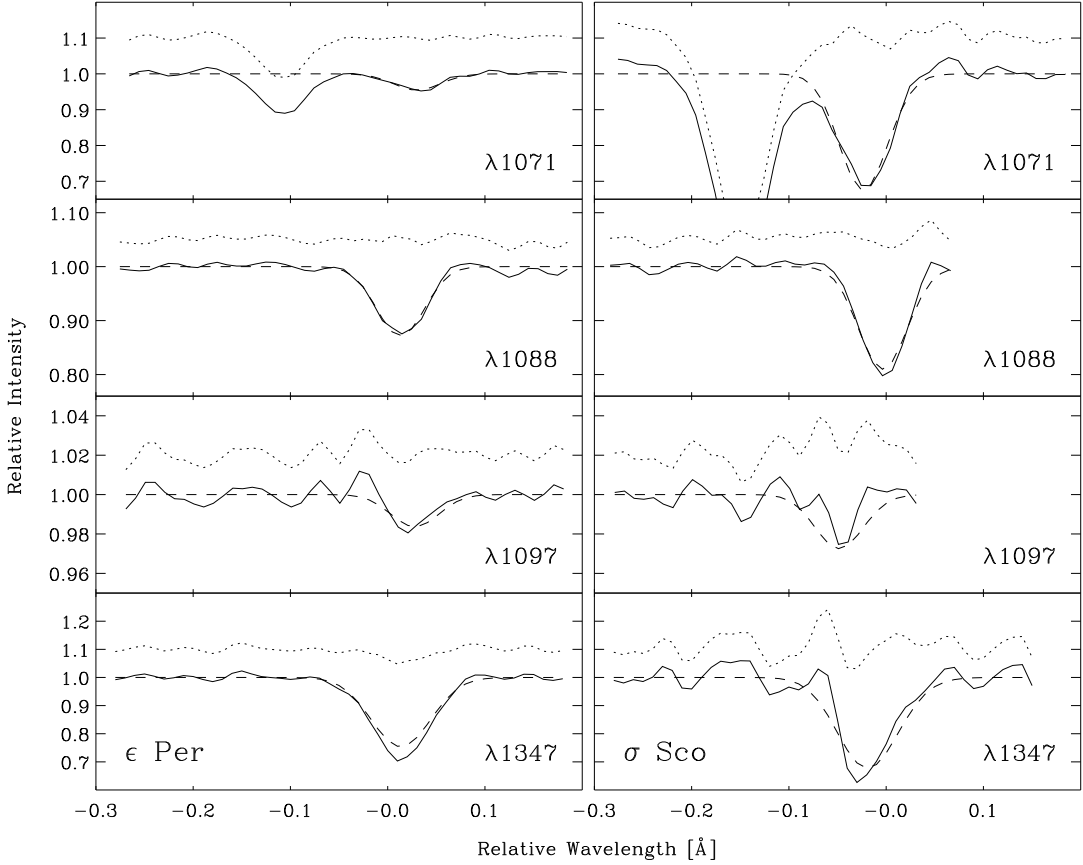}
\vspace{0.3in}
\caption{Sample spectra of Cl~{\small II} and Cl~{\small I} lines toward 
$\epsilon$ Per and $\sigma$ Sco.  Vertical scales from panel to panel vary.  
Absorption from H$_2$ is seen to the blue of Cl~{\small II} $\lambda$1071.  
Data are shown as solid lines, fits based on the weighted average 
Cl~{\small I} column densities by dashed lines, and residuals by 
dotted lines.  The slight offset in relative wavelength results from the 
accuracy of the original wavelength scale.}
\end{figure}

\clearpage

\begin{figure}
\hspace{1.5in}
\includegraphics[scale=0.67, angle=90]{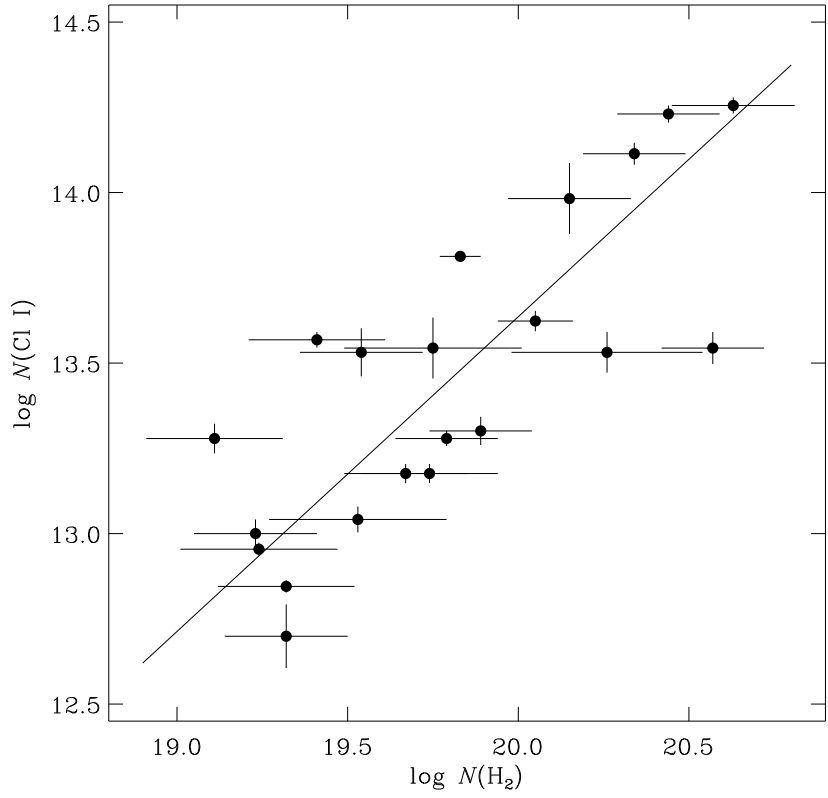}
\vspace{0.3in}
\caption{Log $N$(Cl~{\small I}) vs. log $N$(H$_2$).  Displayed uncertainities 
are 1 $\sigma$.  The linear fit is based on treating uncertainties in x and y 
independently.  The resulting slope is $0.92\pm0.19$.}
\end{figure}

\clearpage

\begin{figure}
\hspace{1.5in}
\includegraphics[scale=0.67, angle=90]{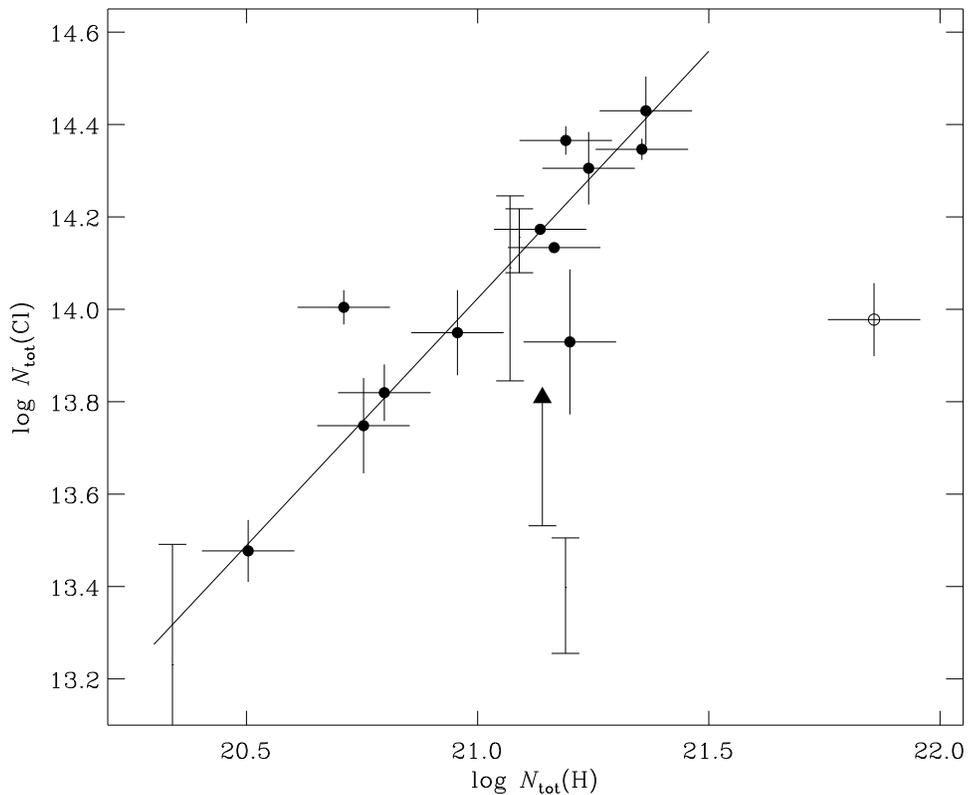}
\vspace{0.3in}
\caption{Log $N_{tot}$(Cl) vs. log $N_{tot}$(H).  Uncertainties are 1 
$\sigma$.  Vertical lines with endcaps indicate directions with upper 
limits to $N$(Cl~{\small II}), while the vertical arrow represents gas 
toward 67 Oph for which no Cl~{\small II} spectrum is available.  The 
result for $\rho$ Oph is shown as an open circle.  Sight lines without 
measures of $N$(H~{\small I}) lie at least 1 dex to the left of the 
data shown here.  A BCES linear fit 
to the filled circles yields a slope of $1.07\pm0.14$.}
\end{figure}
\end{document}